\documentstyle[12pt,epsfig]{article}

\newlength{\dinwidth}
\newlength{\dinmargin}
\def\lapproxeq{\lower .7ex\hbox{$\;\stackrel{\textstyle
<}{\sim}\;$}}
\def\gapproxeq{\lower .7ex\hbox{$\;\stackrel{\textstyle
>}{\sim}\;$}}

\def\beq{\begin{equation}}
\def\eeq{\end{equation}}
\def\bea{\begin{eqnarray}}
\def\eea{\end{eqnarray}}

\def\GeV{\rm GeV}
\def\msbar{\overline{\rm MS}}

\begin{document}
\begin{flushright}
IPPP/07/23 \\
DCPT/07/46 \\
11th July 2007 \\

\end{flushright}

\vspace*{0.3cm}

\begin{center}
{\Large \bf Update of Parton Distributions at NNLO }

\vspace*{0.8cm}
\textsc{A.D. Martin$^a$, W.J. Stirling$^a$,
R.S. Thorne$^{b,}$\footnote{Royal Society University Research Fellow} and
G. Watt$^b$} \\

\vspace*{0.5cm} $^a$ Institute for Particle Physics Phenomenology,
University of Durham, DH1 3LE, UK \\
$^b$ Department of Physics and Astronomy, University College London,
WC1E 6BT, UK
\end{center}

\vspace*{0.5cm}

\begin{abstract}
We present a new set of parton distributions obtained at NNLO. 
These differ from the previous sets available at NNLO due to improvements 
in the theoretical treatment. In particular we include a full treatment 
of heavy flavours in the region near the quark mass. 
In this way, an essentially complete set of NNLO partons 
is presented for the first time. The improved treatment leads to a 
significant change in the gluon and heavy quark distributions, and a larger 
value of the QCD coupling at NNLO, $\alpha_S(M_Z^2)=0.1191\pm 
0.002({\rm expt.}) \pm 0.003({\rm theory})$. Indirectly this also leads to a 
change in the light partons at small $x$ and modifications of our predictions 
for $W$ and $Z$ production at the LHC. As well as the best-fit set of 
partons, we also provide 30 additional sets representing the uncertainties 
of the partons obtained using the Hessian approach.   
\end{abstract}

The highest level of precision so far achieved for cross sections calculated 
in fixed-order perturbative QCD is next-to-next-to-leading order (NNLO).
The cross sections are known at NNLO for many processes involving incoming
protons (or antiprotons), i.e.~coefficient functions exist for deep-inelastic 
scattering \cite{CF}, for hadro-production of electroweak particles 
such as $W$ and 
$Z$ bosons \cite{DY} (as a function of rapidity \cite{NNLODY}), 
and even Higgs bosons \cite{HiggsNNLO,Higgsrap} 
(in both the standard and supersymmetric models). 
The appropriate parton distributions to use with these hard cross sections are 
those which have been evolved using NNLO evolution kernels. Approximations to 
these \cite{VV12} based on exact moments \cite{moments} 
and small-$x$ limits \cite{S47} have been 
available for a number of years, and correspondingly
approximate NNLO parton distributions have been extracted from 
global fits \cite{MRSTapproxNNLO,MRSTerror2}, 
or fits to DIS data \cite{ALapproxNNLO}. 
The complete calculation of the splitting functions was 
published in 2004 \cite{NNLOs} 
(the results being numerically very similar to the previous estimates)
and this allowed improved NNLO parton distributions to be obtained
\cite{MRST04, ALexact}. 
Nevertheless, some additional, sometimes unmentioned, 
approximations still exist in the published sets 
of NNLO partons.  

\begin{figure}[ht]
\vspace{-0.6cm}
\centerline{\epsfxsize=3in\epsfbox{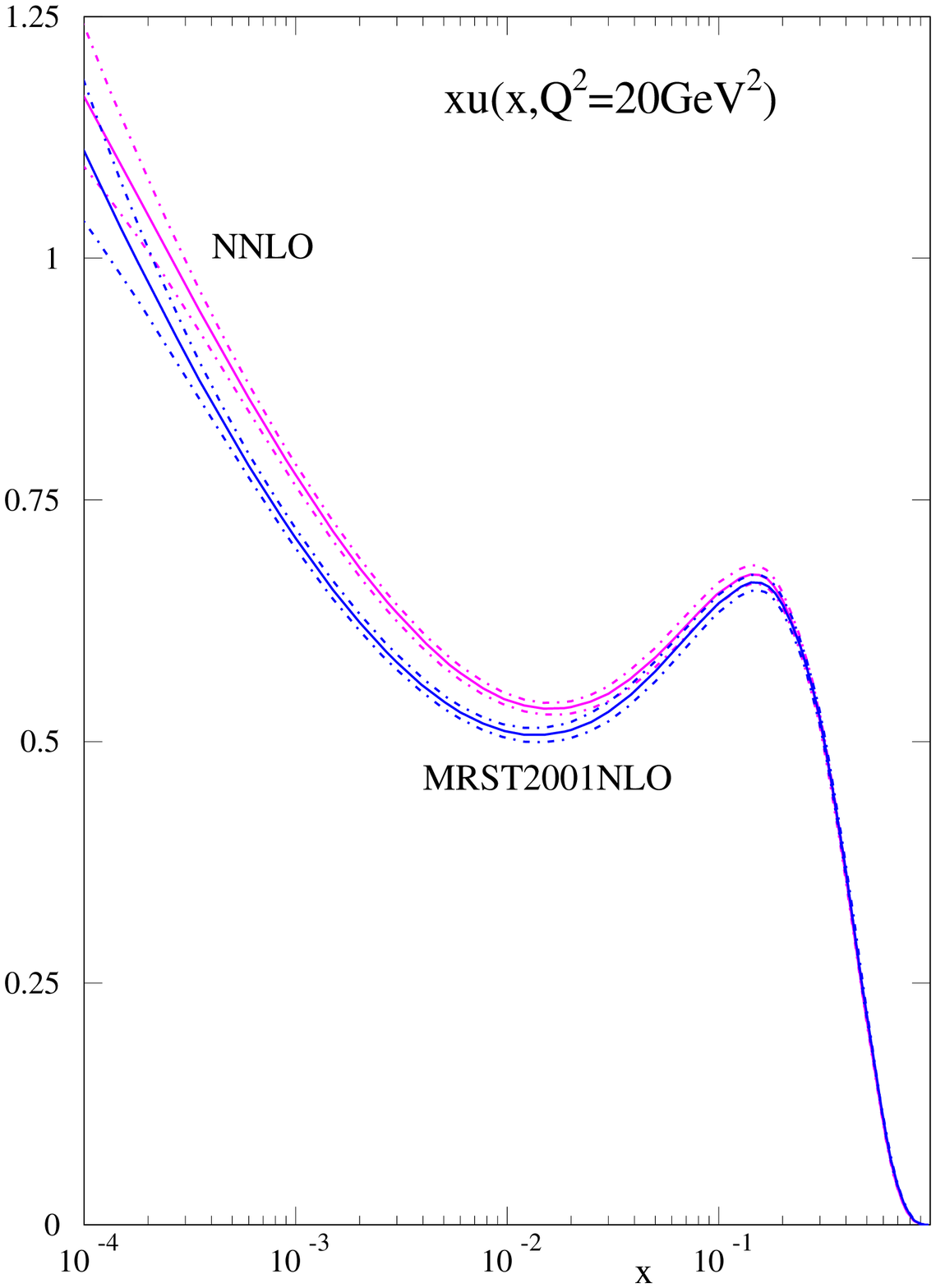}
\hspace{0.8cm}\epsfxsize=3in\epsfbox{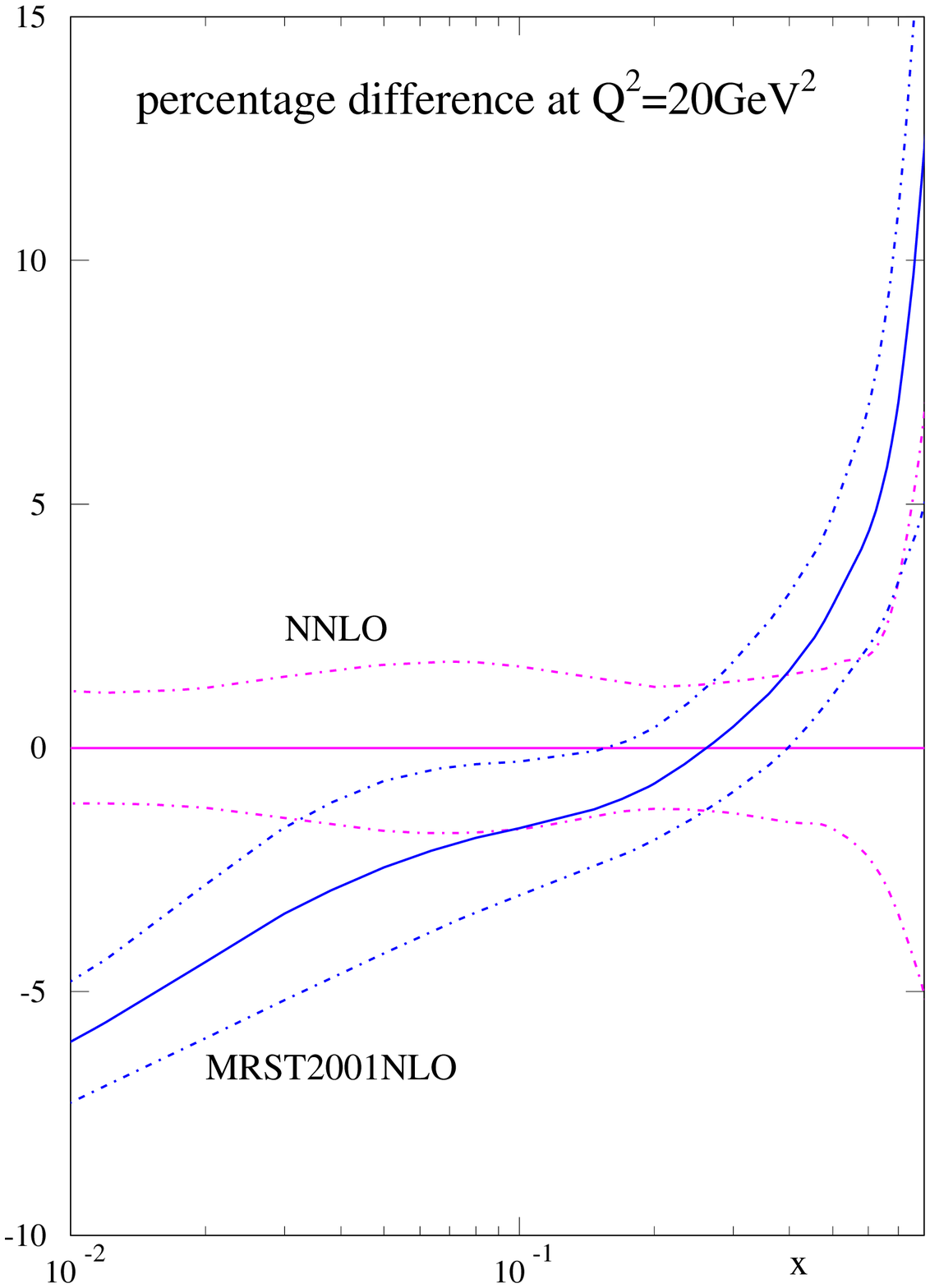}}   
\vspace{-0.1cm}
\caption{Comparison of the NLO up distribution with the NNLO up distribution,
concentrating on small $x$ (left) and high $x$ (right).
  \label{fig1}}
\vspace{-0.1cm}
\end{figure}


One of the most significant of these is the treatment of heavy flavour. 
At NNLO the partons become discontinuous at flavour transition points 
(conventionally chosen to be at $\mu^2=m_H^2$, with $H=c,b$), due to the 
non-vanishing of transition matrix elements \cite{buza}
if the flavour number is allowed 
to vary. There is no reason why any parton should be exactly zero at a 
particular scale, and it is an accident of the $\msbar$ renormalisation and 
factorisation scheme that the  structure of the perturbative
heavy flavour is so simple that one can evolve from zero at a given scale 
even at NLO. This is not true, for example, in the MS renormalisation scheme
or even in the DIS factorisation scheme (an all orders heavy flavour 
prescription in this scheme is presented in \cite{White}) where an 
$x$-dependent input at $\mu^2=m_H^2$ is required at NLO.     
These discontinuities in NNLO partons 
have so far been ignored, as explained in the appendices of 
\cite{MRSTapproxNNLO}. 
In principle it is possible to 
avoid this problem of discontinuities at the transition points by simply 
remaining in a fixed flavour number scheme. However, in practice  
the heavy flavour coefficient functions appropriate for this 
choice are only known to NLO so this is not an option.
A complete NNLO set of partons must have a properly defined treatment 
of heavy flavour. None of the previous sets satisfy this requirement. 
Also, the NNLO parton distributions should be obtained by 
a comparison to data using cross sections calculated at the same order.  

\begin{figure}[ht]
\vspace{-0.6cm}
\centerline{{\epsfxsize=3in\epsfbox{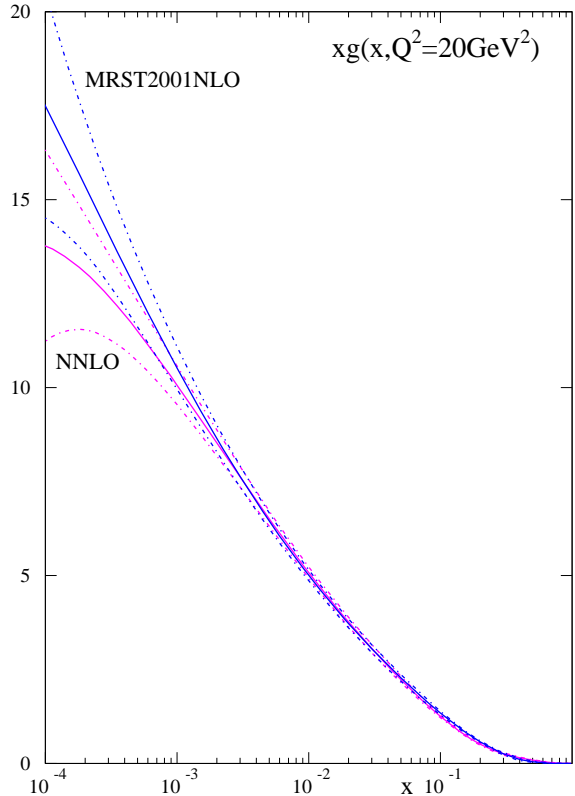}}}   
\vspace{-0.1cm}
\caption{Comparison of the NLO gluon distribution with the NNLO gluon 
distribution, concentrating on small $x$.
  \label{fig3}}
\vspace{-0.1cm}
\end{figure}

\begin{figure}[ht]
\vspace{-0.6cm}
\centerline{{\epsfxsize=3in\epsfbox{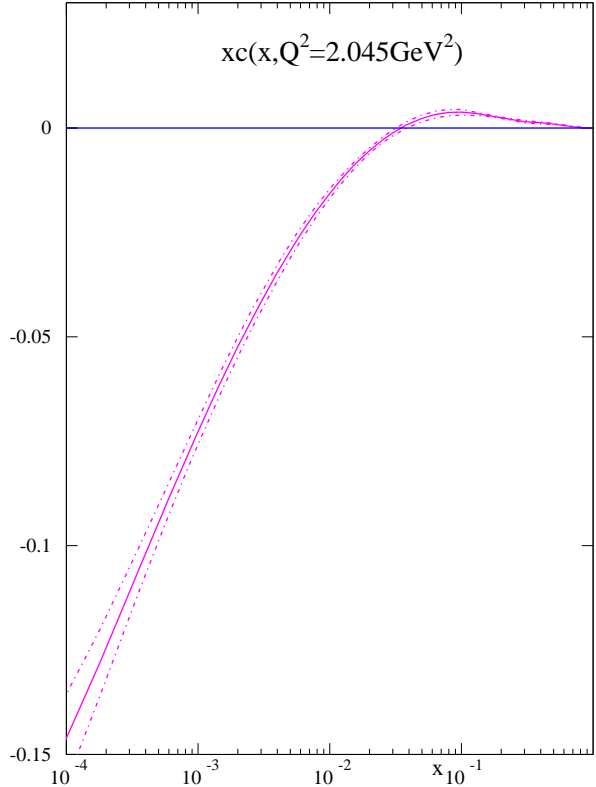}}}   
\vspace{-0.1cm}
\caption{The charm distribution, $xc(x,\mu^2)$,  
at the transition point $\mu^2=m_c^2=2.045\,\GeV^2$. 
The distribution, which is zero for 
$\mu^2<m_c^2$, turns on, at NNLO in the $\msbar$ scheme, 
with a non-zero value at $\mu^2=m_c^2$ given by the curve above. 
(The discontinuity in the charm distribution,
and other parton densities, at $\mu^2=m_c^2$ is compensated by
discontinuities in the coefficient functions such that structure functions are
continuous up to tiny N$^3$LO contributions \cite{nnlovfns}.) 
  \label{fig4}}
\vspace{-0.1cm}
\end{figure}

\begin{figure}[ht]
\vspace{-0.6cm}
\centerline{{\epsfxsize=3in\epsfbox{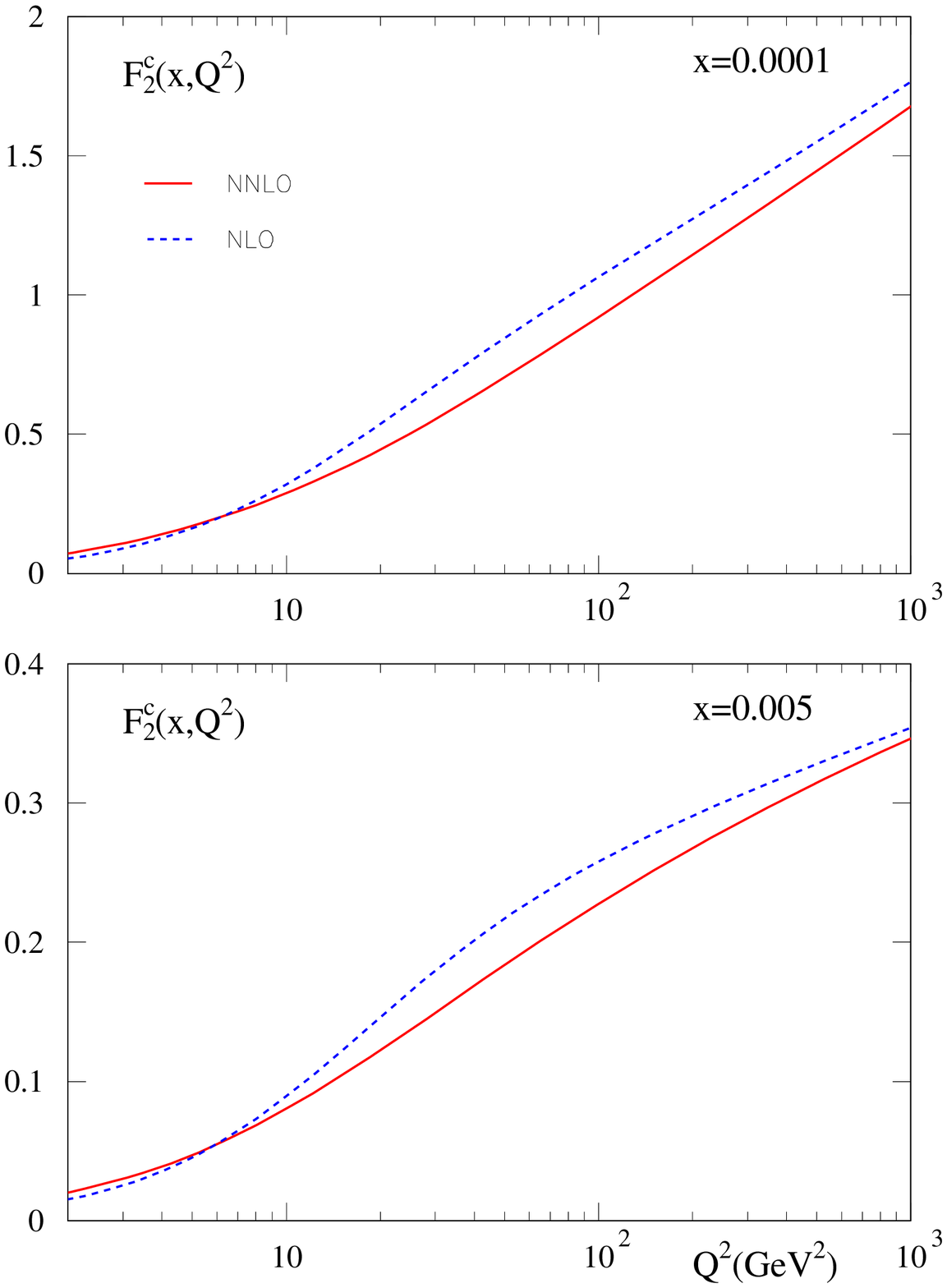}}}   
\vspace{-0.1cm}
\caption{The charm contribution to $F_2(x,Q^2)$ at NLO and NNLO as a 
function of $Q^2$.
  \label{fig5}}
\vspace{-0.1cm}
\end{figure} 

These considerations make it necessary to update the NNLO parton 
distributions. Compared to the previous version, MRST2004 \cite{MRST04}, 
there are two main changes in the theory: an implementation of 
a new variable flavour number scheme (VFNS) for the heavy quark 
$(c,b)$ flavours \cite{nnlovfns} (based on the NLO schemes in 
\cite{acotchi} and \cite{trvfns}) which maintains the continuity 
(up to N$^3$LO corrections which are very small) of both neutral and 
charged current structure functions by 
introducing discontinuities in coefficient functions which counter 
those in the parton distributions; and the inclusion of 
NNLO corrections \cite{NNLODY} to the Drell--Yan cross sections. 
The most important change compared to the previous 
NNLO partons, which already used the exact splitting 
functions \cite{NNLOs},  
is the new VFNS which leads to a significant change in the gluon
and heavy quark distributions. Moreover, because we knew that the previous 
NNLO treatments were 
approximate, we did not provide uncertainties along with the central values. 
Now that the NNLO procedure  
is essentially complete we rectify this.\footnote{Some of the 
most important features of these new 
distributions have been highlighted in \cite{DIS06}.}
The size of the uncertainties 
due to the experimental errors on the data fitted, which is obtained using 
the Hessian approach
\cite{CTEQHes}, is similar to that at NLO \cite{MRSTerror1}. 
There is more work to do in order to estimate the theoretical uncertainty, 
which is certainly important in some regions \cite{MRSTerror2}.

\begin{figure}[ht] 
\vspace{-0.5cm}
\centerline{\hspace{-1.3cm}\epsfxsize=3in\epsfbox{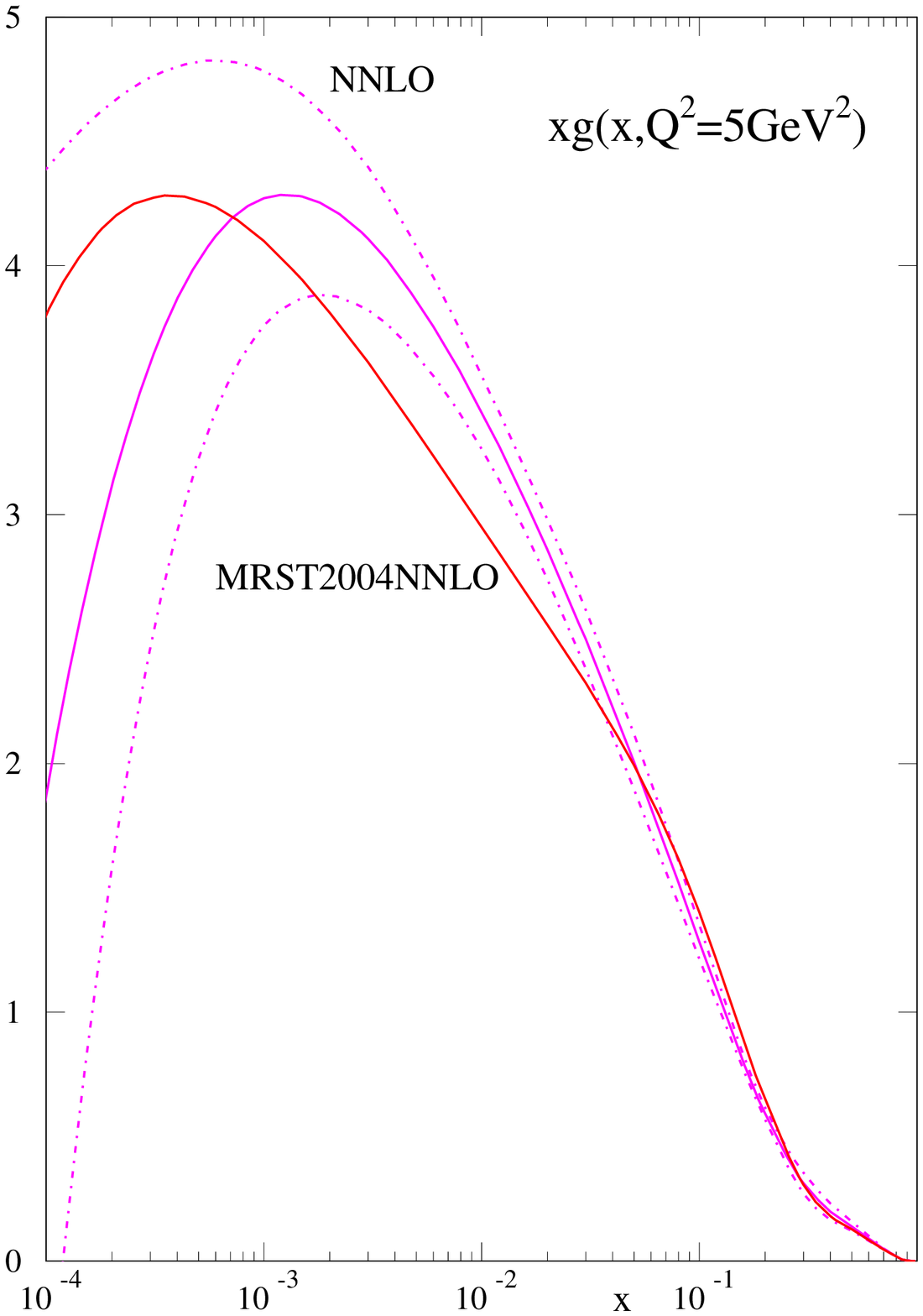}
\hspace{2.3cm}\epsfxsize=2.74in\epsfbox{ratio2006to2004.eps}}   
\vspace{-0.1cm}
\caption{Comparison of the NNLO gluon distribution (together with its 
uncertainty) with the previous approximate NNLO distribution at 
$Q^2=5~\GeV^2$ (left), and the ratio at $Q^2=10^4~ \GeV^2$
for both the gluon and the up quark (right).
  \label{fig6}}
\vspace{-0.1cm}
\end{figure}

\begin{figure}[ht]
\vspace{-0.6cm}
\centerline{{\epsfxsize=3in\epsfbox{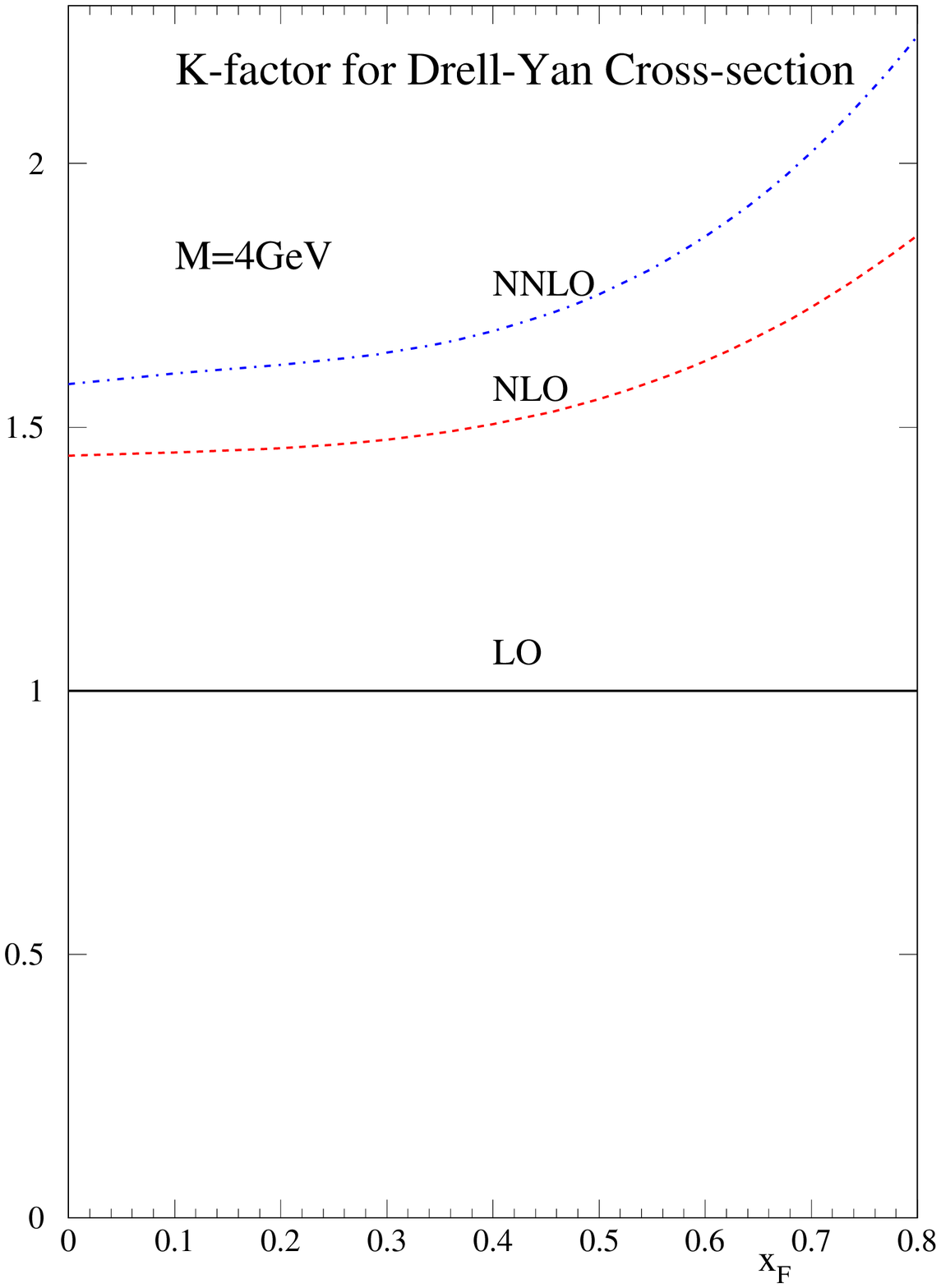}}}   
\vspace{-0.1cm}
\caption{The $K$-factors for Drell--Yan production using LO, NLO and NNLO
partonic cross sections and a fixed set of NNLO parton distributions.
  \label{fig7}}
\vspace{-0.1cm}
\end{figure}

The procedure that we use to obtain the partons, and their uncertainties, is 
very 
similar to previous analyses, with the data used in the fit being 
essentially the 
same as in the MRST2004 analysis. We provide 15 different eigenvector sets 
of partons, with an ``up'' and ``down'' set for each. Each of the 
30 parton sets 
corresponds to a $\Delta \chi^2$ of 50 compared to the best fit, this 
corresponding to an approximate $90\%$ confidence-level for the uncertainty 
of the partons.\footnote{Note that it does not always correspond precisely to 
a rescaling of the uncertainty for $\Delta \chi^2=1$ by a factor of 
$\sqrt{50}$, because the increase in $\chi^2$ about the minimum is not 
completely quadratic for every eigenvector.} 
These partons can be used to estimate the uncertainty of 
physical quantities in exactly the manner explained for the NLO partons in 
\cite{MRSTerror1}. 

\begin{figure}[ht]
\centerline{\hspace{-0.7cm}\epsfxsize=3in\epsfbox{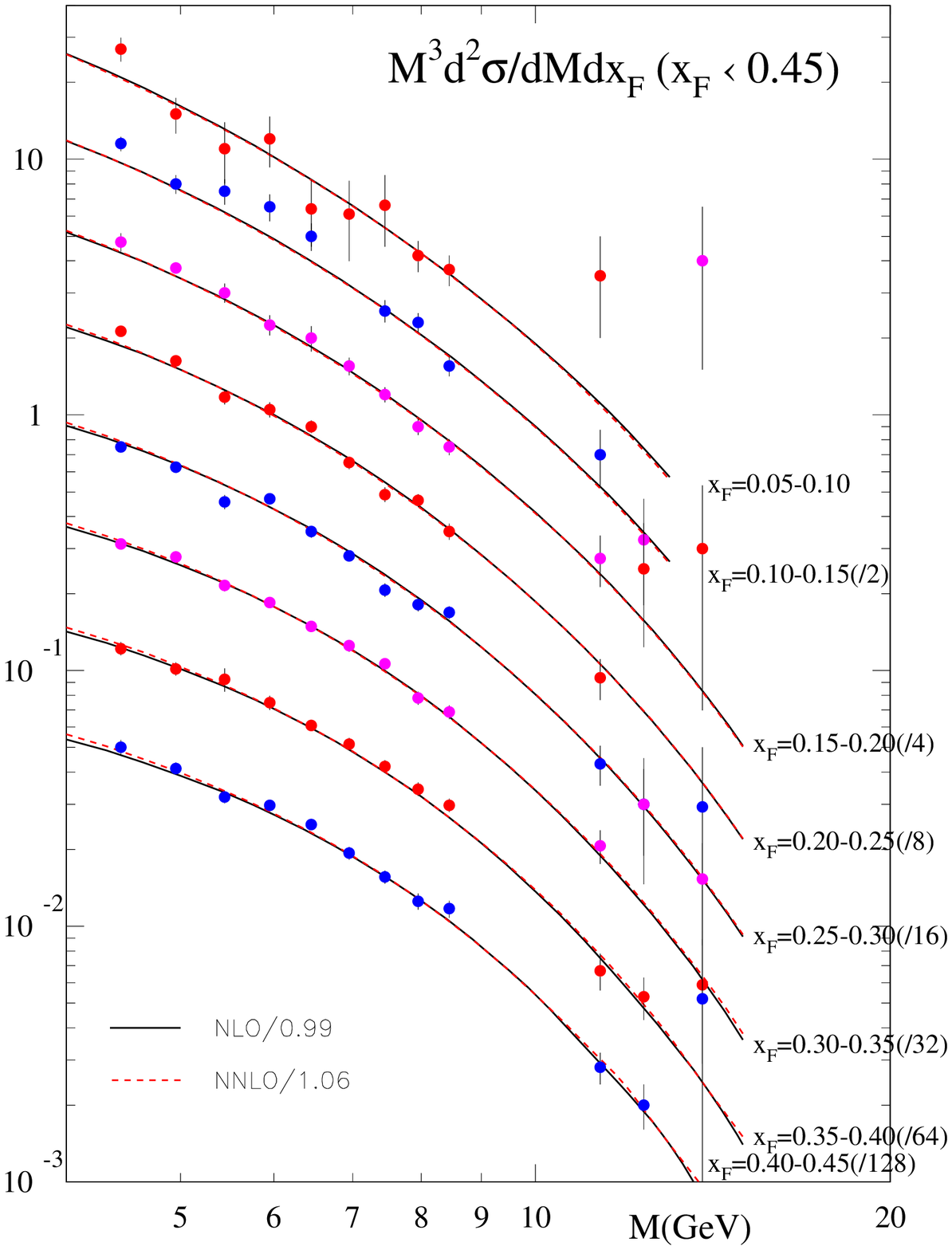}
\hspace{0.5cm}\epsfxsize=3in\epsfbox{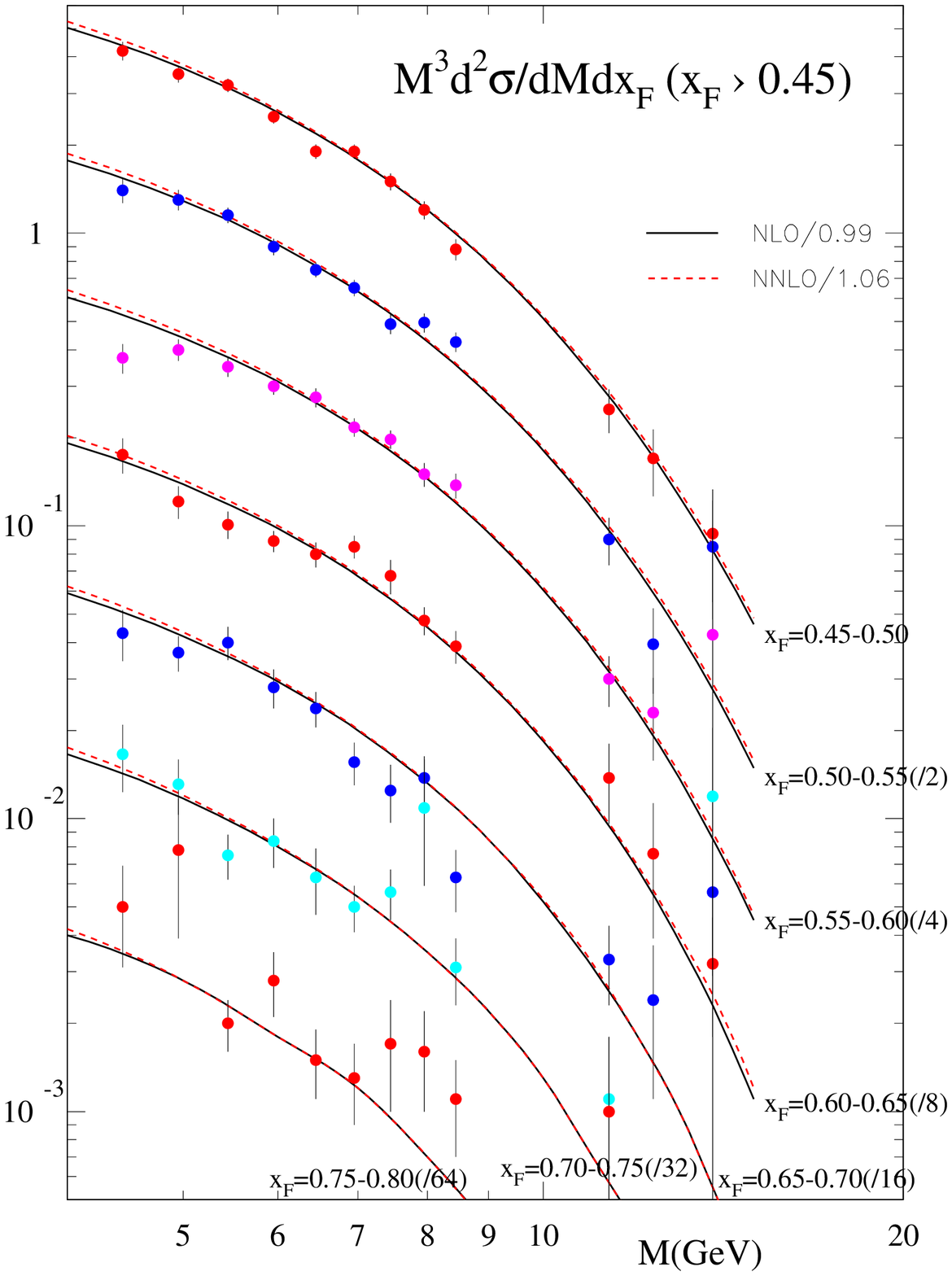}}   
\vspace{-0.1cm}
\caption{Comparison of the NLO and NNLO Drell--Yan cross sections with the 
data. 
  \label{fig8}}
\vspace{-0.1cm}
\end{figure}

The central values for 
the light quarks are similar to the previous NNLO analysis, though 
the evolution with $Q^2$ is a little different in order to obtain the best 
fit with the different heavy flavour procedure. This leads to the most 
marked change being at small $x$ and high $Q^2$, and we will return to this.
At lower $Q^2$ it is most illuminating to instead 
examine the difference between the NNLO quarks and the quarks at 
NLO \cite{MRSTerror1}.   
The change in the up quark distribution when going from NLO to NNLO 
is shown in 
Fig.~\ref{fig1}. At small $x$ the effect of the coefficient functions,
particularly $C^{(2)}_{2,g}(x)$, is important and
the difference between  the NLO and the NNLO distribution is 
greater than the uncertainty in each.
At large $x$ the coefficient functions are again important, i.e.
$$
C^{(2)}_{2,q}(x)\sim \biggl(\frac{\ln^3(1-x)}{(1-x)}\biggr)_+
$$ 
and the 
difference between NLO and NNLO is again larger than the uncertainty 
in each as seen in Fig.~\ref{fig1}. 
At small $x$ the effect of the splitting functions is also important, 
particularly due to $P^{(2)}_{qg}(x)$, which has a
positive $\ln(1/x)/x$ contribution.
This affects the gluon distribution via the fit to 
$dF_2(x,Q^2)/d\ln Q^2$,
and the NNLO gluon is  smaller at very low $x$ than the NLO gluon, as shown 
in Fig.~\ref{fig3}.

The major change in the partons comes about due to the improved treatment 
of the 
heavy quarks, $H=c,b$. As before, we assume that the heavy quark 
distributions are generated by perturbative evolution, i.e.~by 
$g, \Sigma \to H{\bar H}$ transitions, where $\Sigma$ is the singlet light 
quark distribution. 
At NNLO, heavy flavour no longer evolves from zero at $\mu^2=m_H^2$. Rather 
the distributions, $H=c,b$, have an input value given by the convolution 
$$
(H+\bar H)(x,m_H^2) = A^{(2)}_{Hg}(m_H^2)\otimes g(m_H^2)+ 
A^{(2)}_{Hq}(m_H^2)\otimes \Sigma(m_H^2),
$$
where, in practice, the heavy flavour distribution 
starts from a negative value at low $x$, since the main contribution is 
from the matrix element at small $x$ which is 
$$
A^{(2)}_{Hg}(x, \mu^2=m_H^2) \to \biggl(\frac{\alpha_S(m_H^2)}
{4\pi}\biggl)^2\biggl(
\frac{40}{3}-\frac{8\pi^2}{3}\biggr)\frac{1}{x},
$$
plus less singular contributions, and the singlet matrix element 
has the same small-$x$ behaviour up to a colour factor of $4/9$. 
These negative small-$x$ matrix elements are combined with the positive 
high- and moderate-$x$ parton distributions in the convolution to give 
negative heavy flavour contributions.  The 
resulting discontinuity at $\mu^2=m_H^2$ in the heavy 
flavour distributions is by no means 
insignificant, see Fig.~\ref{fig4}. 
The discontinuities make the use of a general-mass variable-flavour scheme 
essential, otherwise the structure functions would have sizable 
discontinuities at NNLO \cite{nnlovfns}.\footnote{One could choose the 
transition point away from $\mu^2=m_H^2$, but there is little sensitivity 
to this choice since the $\ln(\mu^2/m_H^2)$-dependent terms in the matrix 
elements account for evolution (without a complete resummation, of course).  
If one made the transition at $\mu^2 \sim 2m_H^2$ the input for the heavy 
partons would be  smaller, although it would not be exactly zero 
anywhere. This would 
be more complicated to implement and the discontinuities in the other partons 
would change as well.} Alternatively, ignoring the 
discontinuities leads to large errors in the heavy flavour partons.   
The discontinuities in the light partons are much less significant, being at 
most a few percent for the gluon, the change quickly being overshadowed 
by evolution, and are very small indeed for light quarks.  
The increased evolution from the  
NNLO splitting function allows the NNLO charm distribution to catch 
up partially 
with respect to that at NLO, which starts from zero at $m_c^2$,
but it always lags a little behind at higher $Q^2$. 
(Similarly the correct NNLO charm 
distribution is smaller than the approximate MRST2004 NNLO
distribution which turned on from zero.) However, the added effect of the 
coefficient function at NNLO (estimated at low $Q^2$ in \cite{nnlovfns}) 
is to raise the NNLO value of $F_2^c(x,Q^2)$ above the NLO value. The 
evolution of $F_2^c(x,Q^2)$ at NNLO compared to NLO is shown in 
Fig.~\ref{fig5}. The increase at low $Q^2$ and the decrease which 
persists to high $Q^2$, means unambiguously that  
$F_2^c(x,Q^2)$ tends to be flatter in $Q^2$ at NNLO than NLO. Note that 
this comparison is made after a refit to data so that the corresponding 
evolution for light quarks must be such as to fit the HERA inclusive data on 
$F_2(x,Q^2)$. In detail this means that 
the light quarks at NNLO now actually need to evolve slightly more 
quickly than at NLO (or than in the previous approximate NNLO treatment) 
to make up for the decrease in the evolution for the heavy 
flavour.  This 
correction in the charm procedure then automatically affects the gluon, 
since it has to compensate 
for the change in evolution of $F_2^c(x,Q^2)$.  The comparison with 
the low $Q^2$ MRST2004 NNLO gluon is shown in the left of Fig.~\ref{fig6},
while the right of Fig.~\ref{fig6} shows the ratio after a long evolution
length for both the gluon and the up quark (the effect on the down quark
being much the same).  We see that the change in the gluon, and on the
light quarks at very high $Q^2$,
is greater than the uncertainty in some places. The correct heavy 
flavour treatment is vital.

\begin{figure}[ht] 
\vspace{-0.5cm}
\centerline{\epsfxsize=3in\epsfbox{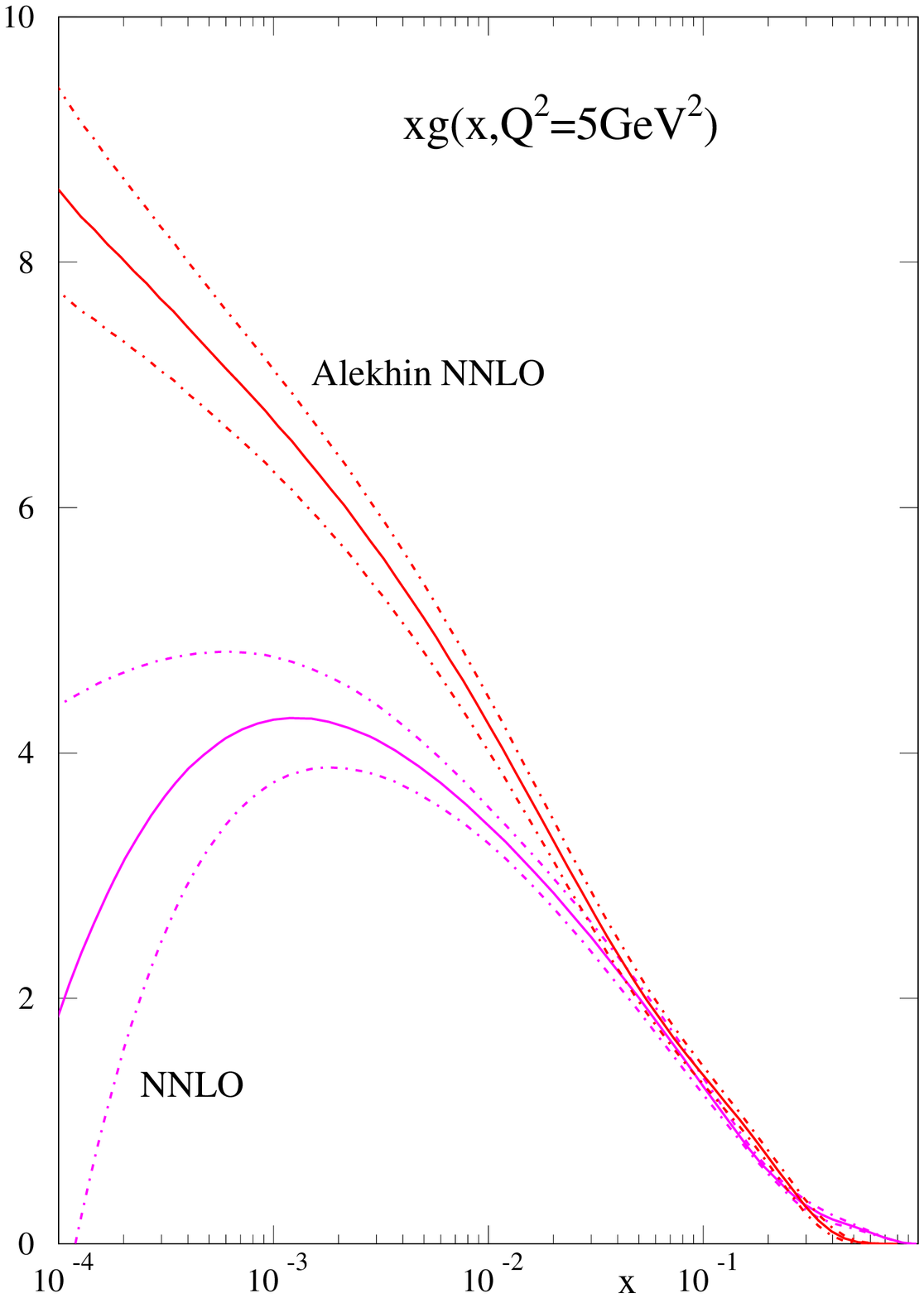}}   
\vspace{-0.1cm}
\caption{Comparison of the NNLO gluon distribution with that of Alekhin.
  \label{fig9}}
\vspace{-0.1cm}
\end{figure}

Our other theoretical improvement at NNLO is also non-trivial. 
The NNLO corrections to the Drell--Yan cross section are significant 
\cite{NNLODY}. 
There is an enhancement at high $x_F=x_1\!-\!x_2$ due to large logarithms,
which is similar 
to  the $\ln(1-x)$ enhancement in structure functions. 
The NLO correction is large and the  NNLO corrections are $10\%$ or 
more, as seen in Fig.~\ref{fig7}.  
The quality of the fit to E866 Drell--Yan production \cite{E866} 
in proton--proton collisions\footnote{We use the data corrected for 
radiative corrections \cite{Reimer}.} is $\chi^2= 223/174$ at  NLO and
$\chi^2= 240/174$ at  NNLO. The
scatter of points is large and a $\chi^2 \sim 220$ is approaching the best 
possible with a smooth distribution. 
The quality of the fit is good, as seen in Fig.~\ref{fig8}. 
It is worse for proton--deuterium data. The positive 
correction at NNLO requires the data normalisation to be $106\%$ 
($99\%$ at NLO), 
there being little freedom since both the sea quarks for 
$x\leq 0.1$ and the valence quarks are already 
well determined by structure function data.
The normalisation uncertainty on the data is $6.5\%$, and a  change of $6\%$
is large but acceptable.\footnote{The NNLO Drell--Yan cross section is also 
used in the latter reference of \cite{ALexact}.} 

The one remaining set of data for which we cannot use NNLO coefficient 
functions is the Tevatron high-$E_T$ inclusive jet data \cite{jets}, 
since the calculation is not 
complete.\footnote{Progress is continually being made in evaluating NNLO 
cross sections of this type \cite{NNLOjets}.} 
However, the systematic uncertainties on these data are of order
$10\%$, which is representative of the size of the NLO corrections, and 
indications \cite{owens} are that the NNLO corrections are 
somewhat less than this (the 
coupling being rather small at these very high scales, of course).    
Whether one includes these data or not can lead to changes in the high-$x$ 
gluon of over $100\%$, so a potential error of probably less than $10\%$ from 
using only the NLO coefficient functions seems trivial in comparison, and we 
keep these data in the fit.    

The quality of the full fit for the set of data used 
is $\chi^2=2406/2287$ at NLO and $\chi^2=2366/2287$ at NNLO. 
The NNLO fit is 
consistently better than the NLO fit. There is most improvement 
in the description of the 
high-$x$ structure function data, where the NNLO coefficient function 
leads to a quicker evolution at low $Q^2$. 
Most DIS data 
sets, and the Tevatron jet data, have a slightly better fit quality at NNLO.
The exception is the E866 Drell--Yan data, as already noted.    
There is a 
tendency for the NNLO $\alpha_S(M_Z^2)$ to increase with
the improved theoretical treatment. An exactly analogous NLO fit yields 
$\alpha_S(M_Z^2)=0.1212$, i.e.~much the same as in \cite{MRST04} since 
the analysis is extremely similar. In comparison, at NNLO 
$\alpha_S(M_Z^2)=0.1191 \pm 0.002({\rm expt.}) \pm 0.003({\rm theory})$, 
whereas in \cite{MRST04} $\alpha_S(M_Z^2)=0.1167\pm 0.002({\rm expt.}) 
\pm 0.003({\rm theory})$. The difference can be traced 
mainly
to the slowing of the heavy flavour evolution using the full treatment, 
leading to an increased coupling as well as a modified gluon distribution 
to compensate. 
Overall, although the fit is generally good, particularly at NNLO, there  is 
some room for improvement, and the data would prefer a little more gluon at 
both high and moderate $x$.

We examine the effect of the change in the NNLO partons on the predictions 
for $W$ and $Z$ production at the LHC and the Tevatron. The results, 
compared to those using the NNLO partons of MRST2004, are shown in 
Table~\ref{tab1},
where we use $B_{l\nu}=0.1068$ and $B_{l^+l^-}=0.033658$. The 
uncertainties on these cross sections are roughly $\pm 2\%$(expt.). 
The results are largely unchanged for the 
Tevatron cross sections, but there is an increase of $6\%$ in the LHC 
cross sections using the updated partons. This is not difficult to understand.
We have already seen the difference in the gluon 
distribution in Fig.~\ref{fig6}.  The larger gluon in the region 
$0.001 \lapproxeq x \lapproxeq 0.1$,
and the larger coupling, are required
to drive the evolution of the light quarks more quickly to match the
small-$x$ HERA data when the new procedure produces a flatter $F_2^c(x,Q^2)$. 
This results in the  larger light quark distributions at high $Q^2$ 
for $x\lapproxeq 0.01$, as seen in the right of Fig.~\ref{fig6}, and 
a 6$\%$ increase in the LHC cross sections, which are dominated by light 
quark--antiquark annihilation with $x \sim 0.006$. The Tevatron probes the
distributions at
 $x \sim 0.05$ resulting in little change in the predictions. 
Ratios of heavy boson cross sections are also essentially unchanged. 
The 6$\%$ change in the predictions of the $W,Z$ rates at the LHC 
should not be regarded as an uncertainty --- it is a consequence of correcting 
an over-simplified treatment of heavy flavours in previous NNLO analyses.
We also note that this change is for different reasons than 
the recent change in predictions for vector boson cross sections at the LHC
observed in going from the CTEQ6.1 to CTEQ6.5 parton distributions 
\cite{CTEQ6.5}. The latter was for NLO partons and came about due to 
the implementation of a general-mass VFNS instead of a zero-mass VFNS. Our
change is due to an improvement in the general-mass VFNS, the most 
important modification being the treatment of parton discontinuities, 
which is a feature which only appears at NNLO.  

\begin{table}
\begin{center}
\begin{tabular}{|l|l|l|}
\hline
&  $B_{l\nu} \cdot \sigma_W ({\rm nb})$ & $B_{l^+l^-}\cdot\sigma_Z ({\rm nb})$
\\
\hline
Tevatron &  2.727 (2.693)   &  0.2534 (0.2518)      \\
LHC  & 21.42 (20.15)  & 2.044 (1.918)     \\
\hline
    \end{tabular}
\caption{Total $W$ and $Z$ cross sections multiplied by leptonic branching
ratios at the Tevatron and the LHC, 
calculated at NNLO using the updated NNLO parton distributions. The 
predictions using the 2004 NNLO sets are shown in brackets.\label{tab1}}
\end{center}
\vspace{-0.5cm}
\end{table}

We compare with the only other publicised NNLO partons available, those of 
Alekhin \cite{ALexact}. 
We have a much larger $\alpha_S(M_Z^2)$,
i.e.~$\alpha_S(M_Z^2)=0.1191\pm 0.002({\rm expt.}) \pm 0.003({\rm theory}) $,
as compared to $0.1143 \pm 0.0014({\rm expt.})$ and $0.1128 
\pm 0.0015({\rm expt.})$ of Refs.~\cite{ALexact}\footnote{It should 
be remembered that there are 
choices in the definition of $\alpha_S$ at a given order, even for a given 
renormalisation scheme. The differences are formally of higher order, but may 
be significant at the level of $1\%$ or so for an extraction of 
$\alpha_S(M_Z^2)$ \cite{coupdiff}. Different parton analyses tend to 
use different definitions. The details of our procedure can be found in 
\cite{Dick} with extension to NNLO.}.
There is not much difference in the high-$x$ valence quarks,
except that explained by the difference in $\alpha_S(M_Z^2)$.   
There are differences in the low-$x$ sea quarks but these are
dominated by differences in the
flavour treatments of $\bar u -\bar d$ and $s(x,Q^2)$. 
The difference between our and Alekhin's gluon distribution at small $x$ is 
seen in Fig.~\ref{fig9}, 
and is much bigger than the uncertainties. This is due to  
the different  heavy flavour treatments, which we do not believe to be 
consistent in \cite{ALexact}\footnote{The fit in 
\cite{ALexact} is performed in the fixed flavour number 
scheme using heavy flavour coefficient functions only up to NLO, and the 
variable flavour scheme partons are generated by evolution of these input 
partons without the inclusion of the discontinuities at 
heavy flavour transition points.}, and which we have already shown to be   
important, as well as to differences in the data fitted and in the 
value of $\alpha_S(M_Z^2)$.
The gluons also differ a great deal at high $x$, where, in our analysis, they 
are determined by the Tevatron jet data \cite{jets} (the 
comparison now being excellent \cite{MRST04}). In \cite{ALexact} 
the gluon is unconstrained here, and 
should presumably have a large uncertainty.  
In the $\msbar$ scheme the gluon is more important for Tevatron jets at high 
$x$ at NNLO than at NLO because the high-$x$ quarks are automatically 
significantly smaller, as seen in Fig.~\ref{fig1}.

In summary, we have new theoretical corrections 
in our global analysis, and have obtained NNLO partons with 
uncertainties for the 
first time. These NNLO partons (denoted MRST2006) can be found at 
{\tt http://durpdg.dur.ac.uk/hepdata/mrs}, where, in order to have a precise 
treatment in the heavy quark transition region, there has been a 
major upgrade of the interpolation code, which now also allows a  
reliable extrapolation outside of the grid (now defined 
for $10^{-6}\leq x \leq 1$
and $1~\GeV^2\leq Q^2 \leq 10^{9}~\GeV^2$) for high $Q^2$ and/or low $x$.
The NNLO fit improves on that at NLO, as well as correcting our previous 
NNLO analysis.
With respect to the latter, 
the value of $\alpha_S(M_Z^2)$ at NNLO moves upwards significantly, and all 
partons change by a significant amount, particularly at small $x$. As we 
have seen, this 
has implications for predictions for processes at the LHC.

For the future, there are more new data to be included:  
neutrino deep-inelastic scattering data from NuTeV \cite{NuTeV} and 
CHORUS \cite{Chorus},
HERA jets \cite{HERAjets}, updated Tevatron high-$E_T$ jets \cite{newcdf}, 
new CDF lepton-asymmetry data in different $E_T$ bins \cite{newleptasym},   
new heavy flavour data from HERA \cite{newcharm}, and a full treatment of 
NuTeV dimuon data \cite{NuTeVdimuon}. There will also soon be averaged 
HERA structure function data \cite{AvHERA}. 
This will lead us to produce fully updated NLO and NNLO partons for 
the LHC complete with uncertainties --- both experimental and theoretical ---
in a relatively short timescale.\footnote{Details of the preliminary MSTW NLO
partons which include all these new data sets can be seen in \cite{MSTW}.}
However, until this major update can be 
finalised, the NNLO partons outlined in this note will serve as by far the 
most theoretically self-consistent, and most stringently constrained set 
currently available at NNLO.      

\vspace{-0.3cm}

\section*{Acknowledgements}
We would 
like to thank Jeppe Andersen for help with the interpolation 
(and extrapolation) code.
RST would like to thank Mandy Cooper-Sarkar and Andreas Vogt 
for discussions, and
the Royal Society for the award of a University Research Fellowship.  GW would 
like to thank the Science and Technology Facilities Council for the award of 
a Responsive Research Associate position. 
The IPPP gratefully acknowledges financial support from the 
Science and Technology Facilities Council.\\


\vspace{-0.85cm}

\begin{thebibliography}{99}

\bibitem{CF} E.B. Zijlstra and W.L. van Neerven, Phys. Lett. {\bf
B272} (1991) 127; ibid {\bf B273} (1991) 476; ibid {\bf B297}
(1992) 377; Nucl. Phys. {\bf B383} (1992) 525;\\
S. Moch, J.A.M. Vermaseren and A. Vogt, Phys. Lett. {\bf B606} (2005) 123;\\
J.A.M. Vermaseren, A. Vogt and S. Moch, Nucl. Phys. {\bf B724} (2005) 3.

\bibitem{DY} R. Hamburg, W.L. van Neerven and T. Matsuura, Nucl. Phys. 
{\bf B359} (1991) 343; Erratum-ibid {\bf B644} (2002) 403. 

\bibitem{NNLODY} C.~Anastasiou, L.J. Dixon, K. Melnikov and F. Petriello,
Phys. Rev. Lett.  {\bf 91} (2003) 182002;
Phys. Rev. {\bf D69} (2004) 094008.

\bibitem{HiggsNNLO} 
R.V. Harlander, W.B. Kilgore, Phys. Rev. Lett. {\bf 88} (2002) 201801;\\
C.~Anastasiou and K. Melnikov, Nucl. Phys. {\bf B646} (2002) 220;\\
V. Ravindran, J. Smith and W.L. van Neerven, Nucl. Phys. {\bf B665} (2003)
325;\\
O. Brein, A. Djouadi and R.V. Harlander, Phys. Lett {\bf B579} (2004) 149;\\
R.V. Harlander and W.B. Kilgore, Phys. Rev. {\bf D68} (2003) 013001.
 
\bibitem{Higgsrap} C.~Anastasiou, K. Melnikov and F. Petriello,
Phys. Rev. Lett.  {\bf 93} (2004) 262002; Nucl. Phys. {\bf B724} (2005)
197.

\bibitem{VV12} W.L. van Neerven and A. Vogt, Nucl. Phys. {\bf
B568} (2000) 263; Nucl. Phys. {\bf B588} (2000) 345; Phys. Lett. {\bf B490}
(2000) 111.

\bibitem{moments} S.A. Larin {\it et al.}, 
Nucl. Phys. {\bf B492} (1997) 338;\\
A. R\'{e}tey and 
J.A.M. Vermaseren, Nucl. Phys. {\bf B604} (2001) 281.

\bibitem{S47} S. Catani and F. Hautmann, Nucl. Phys. {\bf B427}
(1994) 475; \\
V.S. Fadin and L.N. Lipatov, Phys. Lett. {\bf B429} (1998) 127; \\
G. Camici and M. Ciafaloni, Phys. Lett. {\bf B430} (1998) 349.

\bibitem{MRSTapproxNNLO} A.D. Martin, R.G. Roberts, W.J. Stirling and 
R.S. Thorne, Eur. Phys. J. {\bf C18} (2000) 117;
 Phys. Lett. {\bf B531} (2002) 216. 

\bibitem{MRSTerror2}  A.D.~Martin,  R.G. Roberts, W.J. Stirling and 
R.S. Thorne, Eur. Phys. J. {\bf C35} (2004) 325.

\bibitem{ALapproxNNLO} S.~I.~Alekhin, Phys. Rev. {\bf D68} (2003) 014002.

\bibitem{NNLOs} S. Moch, J.A.M. Vermaseren and A. Vogt,  
Nucl. Phys. {\bf B688} (2004) 101; \\
A. Vogt, S. Moch and J.A.M. Vermaseren, 
Nucl. Phys. {\bf B691} (2004) 129.

\bibitem{MRST04}A.D. Martin, R.G. Roberts, W.J. Stirling and R.S. Thorne, 
Phys. Lett. {\bf B604} (2004) 61.

\bibitem{ALexact} 
S.I. Alekhin, JETP Lett. {\bf 82} (2005) 628; \\
S.I. Alekhin, K. Melnikov and F. Petriello, Phys. Rev. {\bf D74} (2006) 
054033.
 
\bibitem{buza}M. Buza {\it et al.}, Nucl. Phys. {\bf B472} (1996) 611;\\
M. Buza {\it et al.}, Eur. Phys. J. {\bf C1} (1998) 301.

\bibitem{White} C.D. White and R.S. Thorne, Phys. Rev. {\bf D74} (2006) 
014002.

\bibitem{nnlovfns} R.S. Thorne, Phys. Rev. {\bf D73} (2006) 054019.

\bibitem{acotchi} W.K. Tung, S. Kretzer and C. Schmidt, J. Phys.
{\bf G28} (2002) 983; \\
S. Kretzer, H.L. Lai, F.I. Olness and W.K. Tung, Phys. Rev.
{\bf D69} (2004) 114005.

\bibitem{trvfns} R.S.~Thorne and R.G.~Roberts, Phys.~Lett. {\bf B421} 
(1998) 303; Phys. Rev. {\bf D57} (1998) 6871.

\bibitem{DIS06} R.S. Thorne, A.D. Martin and W.J. Stirling, proceedings of
``14th International Workshop on Deep Inelastic Scattering (DIS 2006)'', 
Tsukuba, Japan, 20--24 Apr 2006, pp 81--84,
{\tt hep-ph/0606244}.

\bibitem{CTEQHes}CTEQ Collaboration:
J. Pumplin {\it et~al.}, Phys. Rev. {\bf D65} (2002) 014013.

\bibitem{MRSTerror1}
 A.D.~Martin, R.G. Roberts, W.J. Stirling and R.S. Thorne,
Eur. Phys. J. {\bf C28} (2003) 455.

\bibitem{E866}NuSea Collaboration: J.C. Webb {\it et~al.}, 
{\tt hep-ex/0302019};\\
J.C. Webb, Ph.~D. thesis, {\tt hep-ex/0301031}.

\bibitem{Reimer} P.E. Reimer, private communication.

\bibitem{jets} D{\O} Collaboration: B. Abbott {\it et al.}, Phys. Rev. Lett.
{\bf 86} (2001) 1707; \\
CDF Collaboration: T. Affolder {\it et~al.}, Phys. Rev.
{\bf D64} (2001) 032001.

\bibitem{NNLOjets} T. Gehrmann,
 ``Status of NNLO Jet Calculations'',
 in ``Proceedings of FRIF workshop on first principles non-perturbative QCD 
of hadron jets'', LPTHE, Paris, France, 12--14 Jan 2006, pp T004; and 
references therein. 

\bibitem{owens} N. Kidonakis and J.F. Owens, Phys. Rev. {\bf D63} (2001)
054019.

\bibitem{CTEQ6.5} W.K. Tung {\it et al.}, JHEP {\bf 0702} (2007) 053.

\bibitem{coupdiff} J. Huston, J. Pumplin, D. Stump and W.K. Tung, JHEP
{\bf 0506} (2005) 080; \\
M. Whalley, ``LHAPDF'' talk at ``HERA-LHC Workshop Final Meeting'',
DESY, Hamburg, 21--24 March 2005.

\bibitem{Dick} R.G. Roberts, ``The Structure of the Proton: 
Deep Inelastic Scattering'', Cambridge University Press, Cambridge, UK,
1990.

\bibitem{NuTeV}  NuTeV Collaboration: M.~Tzanov {\it et al.}, 
Phys. Rev. {\bf D74} 012008 (2006).

\bibitem{Chorus} CHORUS Collaboration: G.~Onengut {\it et al.},
Phys. Lett. {\bf B632} (2006) 65.

\bibitem{HERAjets} H1 Collaboration: C.~Adloff {\it et al.},  
Eur.Phys. J. {\bf C19} (2001) 289;\\   
ZEUS Collaboration: S.~Chekanov {\it et al.}, 
Phys. Lett. {\bf B547} (2002) 164; 
Nucl. Phys. {\bf B765} (2007) 1.

\bibitem{newcdf} CDF Collaboration: A. Abulencia {\it et al.}, 
{\tt hep-ex/0701051}. 

\bibitem{newleptasym} CDF Collaboration: D. Acosta {\it et al.}, Phys. Rev. 
{\bf D71} (2005) 051104. 

\bibitem{newcharm} H1 Collaboration: A. Aktas {\it et al.}, Eur. Phys. J.  
{\bf C40} (2005) 349; ibid {\bf C45} (2006) 23;\\
ZEUS Collaboration: S. Chekanov {\it et al.}, Phys. Rev. {\bf D69} (2004) 
012004.  

\bibitem{NuTeVdimuon}   NuTeV Collaboration: M.~Goncharov {\it et al.},
  Phys. Rev. {\bf D64} (2001) 112006. 

\bibitem{AvHERA} S. Glazov, proceedings of ``13th International Workshop on
Deep Inelastic Scattering (DIS 2005)'', Madison, Wisconsin,
27 April--1 May 2005, p.~237;\\
M. Dittmar {\it et al.},  ``Working Group I: Parton distributions: Summary 
report for the HERA-LHC Workshop Proceedings'', p.~54, {\tt hep-ph/0511119},
{\tt hep-ph/0601012}.

\bibitem{MSTW} R.S. Thorne, A.D. Martin, W.J. Stirling and G. Watt, 
proceedings of ``15th International Workshop on Deep Inelastic Scattering
(DIS 2007)'', Munich, 16--20 April 2007,
{\tt arXiv:0706.0456} [hep-ph].

\end{thebibliography}
\end{document}